\documentclass[camera]{jpaper}

\usepackage{multirow}
\usepackage{mathtools}
\usepackage{xspace}
\usepackage{textcomp}
\usepackage{listings}
\usepackage[plain]{algorithm}  % the floater wrapper for algorithmic
\usepackage[noend]{algpseudocode}
\usepackage{algorithmicx}
\floatname{algorithm}{Template}
\usepackage{xfrac}
\usepackage{subscript}

\usepackage{dblfloatfix}
\usepackage{appendix}
\usepackage{authblk}

\usepackage{float}
\usepackage{booktabs}
\usepackage{placeins}
\usepackage{balance}
\usepackage{gensymb}
\usepackage[absolute]{textpos}
\usepackage[nocompress]{cite}
\usepackage{array}

\makeatletter
\let\MYcaption\@makecaption
\makeatother

\usepackage[font=footnotesize]{subcaption}

\makeatletter
\let\@makecaption\MYcaption
\makeatother

\usepackage{fixltx2e}
\usepackage{dblfloatfix}
\usepackage[nolessnomore, italic]{mathastext}
\usepackage[T1]{fontenc}
\usepackage[usenames,dvipsnames,svgnames,table]{xcolor}
\usepackage[normalem]{ulem}
\usepackage{enumitem}
\usepackage{setspace}
\usepackage{indentfirst}
\usepackage{footmisc}
\usepackage{fancyhdr}
\usepackage[us,12hr]{datetime}
\usepackage[keeplastbox]{flushend}
\usepackage[hidelinks]{hyperref}

\newif\ifcameraready
\camerareadytrue
\newcommand{\versionnum}[0]{5.1}

\ifcameraready
\else
\fi

\ifcameraready
  \newcommand{\todo}[1][]{}
  \newcommand{\ch}[0]{}
\else
  \newcommand{\todo}[1][]{\textbf{\fcolorbox{black}{red}{\color{white}{TODO}}} \underline{$\overline{\hbox{\emph{#1}}}$}}
  \newcommand{\ch}[1]{{\color{BrickRed} #1}}
\fi

\usepackage{ifthen}
\usepackage{calc}
\usepackage{pifont}
\usepackage{color}
\usepackage{xcolor}
\usepackage{fancyhdr}
\usepackage{tikz}

\newdimen\origiwspc%
\origiwspc=\fontdimen2\font% original inter word space

\newcommand{\squishlist} {
	\begin{list}{$\bullet$} {
		\setlength{\itemsep}{-2pt}
		\setlength{\parsep}{2pt}
		\setlength{\topsep}{0pt}
		\setlength{\partopsep}{0pt}
		\setlength{\leftmargin}{1.0em}
		\setlength{\labelwidth}{1em}
		\setlength{\labelsep}{0.5em}
	}
}
\newcommand{\squishend} {
	\end{list}
}

\newcounter{hours}
\newcounter{minutes}

\newboolean{timeofmake}
\setboolean{timeofmake}{false}

\RequirePackage{enumerate}
\RequirePackage{enumitem}
\setlist{leftmargin=*, topsep=0pt, partopsep=0pt}
\def\HiLi{\leavevmode\rlap{\hbox to \hsize{\color{yellow!50}\leaders\hrule height .8\baselineskip depth .5ex\hfill}}}

\newcommand{\ignore}[1]{}
\newcommand{\new}[0]{}
\newcommand{\newt}[0]{}
\newcommand{\newthree}[0]{}

\newcommand{\fix}[0]{}

\newcommand{\romnum}[1]{{\em (#1)}\xspace}

\newcommand{\dimm}[2]{$\texttt{D}^{#1}_{#2}$\xspace}

\newcounter{observation}
{% begin code
    \par
    \refstepcounter{observation}%
    \textbf{Observation \theobservation:}
    \begin{itshape}%
    }%
    {% end code
    \end{itshape}%
}
\newcommand{\mech}[0]{\mbox{FLY-DRAM}\xspace}
\newcommand{\mechlong}[0]{{Flexible-LatencY DRAM}\xspace}

\newcommand{\act}[0]{\textsc{{activate}}\xspace}

\newcommand{\crd}[0]{\textsc{{read}}\xspace}
\newcommand{\cwr}[0]{\textsc{{write}}\xspace}
\newcommand{\cpre}[0]{\textsc{{precharge}}\xspace}

\newcommand{\tras}[0]{{tRAS}\xspace}

\newcommand{\trp}[0]{{tRP}\xspace}
\newcommand{\trpe}[0]{{tRP}=}

\newcommand{\trcd}[0]{{tRCD}\xspace}
\newcommand{\trcde}[0]{{tRCD}=}

\newcommand{\tcl}[0]{{tCL}\xspace}

\newcommand{\figref}[1]{Figure~\ref{fig:#1}}
\newcommand{\tabref}[1]{Table~\ref{tab:#1}}

\newcommand{\secref}[1]{Section~\ref{sec:#1}}
\newcommand{\ssecref}[1]{Section~\ref{ssec:#1}}

\newcommand{\figputHS}[3]{
\begin{figure}[h]
\begin{minipage}{\linewidth}
\begin{center}
\includegraphics[scale=#2]{plots/#1}
\end{center}
\vspace{-0.1in}
\caption{#3 \label{fig:#1}}
\end{minipage}
\end{figure}
}

\newcommand{\figputHSL}[4]{
\begin{figure}[h]
\begin{minipage}{\linewidth}
\begin{center}
\includegraphics[scale=#2]{plots/#1}
\end{center}
\vspace{-0.1in}
\caption{#3 \label{fig:#4}}
\end{minipage}
\end{figure}
}

\title{Flexible-Latency DRAM: Understanding and Exploiting\\ Latency Variation in
  Modern DRAM Chips}

\author{%
{Kevin K. Chang$^{1,2}$}%
\qquad%
{Abhijith Kashyap$^{3,2}$}%
\qquad%
{Hasan Hassan$^{4,2,5}$}%
\qquad%
{Saugata Ghose$^{2}$}%
\vspace{2pt}\\%
{Kevin Hsieh$^{2}$}%
\qquad%
{Donghyuk Lee$^{6,2}$}%
\qquad%
{Tianshi Li$^{2,7}$}%
\qquad%
{Gennady Pekhimenko$^{8,2}$}%
\vspace{2pt}\\%
{Samira Khan$^{9,2}$}% 
\qquad%
{Onur Mutlu$^{4,2}$}% 
}
\affil{
{\it%
$^1$Facebook%
\qquad%
$^2$Carnegie Mellon University%
\qquad%
$^3$NVIDIA%
\qquad%
$^4$ETH Z{\"u}rich%
}\vspace{2pt}\\{\it%
$^5$TOBB University of Economics \& Technology%
\qquad%
$^6$NVIDIA Research%
}\vspace{2pt}\\{\it%
$^7$Peking University%
\qquad%
$^8$University of Toronto%
\qquad%
$^9$University of Virginia%
}}

\date{}

\begin{document}
\maketitle

\begin{abstract}
% !TEX root=../paper.tex

\fix{This article summarizes key results of our work on experimental characterization
and analysis of latency variation  and latency-reliability trade-offs in modern
DRAM chips, which was published in SIGMETRICS 2016~\cite{chang-sigmetrics2016},
and examines the work's significance and future potential.}
\fix{Our work is motivated to reduce the} long DRAM latency, which is a critical
performance bottleneck in current systems.  DRAM access latency is defined by
three fundamental operations that take place within the DRAM cell array: {\em
  (i)}~\emph{activation} of a memory row, which opens the row to perform
accesses; {\em (ii)}~\emph{precharge}, which prepares the cell array for the
next memory access; and {\em (iii)}~\emph{restoration} of the row, which
restores the values of cells in the row that were destroyed due to activation.
There is significant latency variation for each of these operations across the
cells of a single DRAM chip due to irregularity in the manufacturing process.
As a result, some cells are \emph{inherently} faster to access, while others are
inherently slower. Unfortunately, existing systems do not exploit this
variation.

The goal of this work is to {\em (i)}~experimentally characterize and
understand the latency variation across cells within a DRAM chip for
these three fundamental DRAM operations, and {\em (ii)}~develop new
mechanisms that exploit our understanding of the latency variation to
reliably improve performance.  To this end, we comprehensively
characterize 240 DRAM chips from three major vendors, and make \ch{six major}
new observations about latency variation within DRAM.  
\ch{Notably, we} find that
{\em (i)}~there is large latency variation across the cells for each
of the three operations; {\em (ii)}~variation characteristics exhibit
significant spatial locality: slower cells are clustered in certain
regions of a DRAM chip; and {\em (iii)}~the three fundamental
operations exhibit different reliability characteristics when the
latency of each operation is reduced.

Based on our observations, we propose \mechlong (\mech), a mechanism that
exploits latency variation across DRAM cells within a DRAM chip to improve
system performance.  The key idea of \mech is to exploit the spatial locality of
slower cells within DRAM, and access the faster DRAM regions with reduced
latencies for the fundamental operations. Our evaluations show that \mech
improves the performance of a wide range of applications by 13.3\%, 17.6\%, and
19.5\%, on average, for each of the three different vendors' real DRAM chips, in
a \new{simulated} 8-core system.

\fix{We have open sourced the data from our research online. We hope the
characterization and analysis we provide opens up new research directions for
both researchers and practitioners in computer architecture and systems.}

\end{abstract}

% !TEX root=../paper.tex
\section{Introduction}

Over the past few decades, the long latency of memory has been a critical
bottleneck in system performance. Increasing core counts, the emergence of more
data-intensive and latency-critical applications, and increasingly limited
bandwidth in the memory system are together leading to higher memory latency.
Thus, low-latency memory operation is now even more important to improving
overall system performance\new{~\cite{wilkes-sigarch2001, mckee-cf2004,
    dean-cacm2013, kanev-isca2015, superfri, mutlu-imw2013, mutlu-date2017}}.

The latency of a memory request is predominantly defined by the timings of three
fundamental operations: (1)~\emph{activation}, which ``opens'' a row of DRAM
cells to access stored data, (2)~\emph{precharge}, which ``closes'' an activated
row, and (3)~\emph{restoration}, which restores the charge level of each DRAM
cell in a row to prevent data loss.\footnote{We refer the reader to our prior
 works\fix{~\cite{chang-sigmetrics2016, kim-isca2012, lee-hpca2013,
      lee-hpca2015, kim-micro2010, kim-hpca2010,chang-hpca2016, hassan-hpca2016,
      chang-sigmetrics2017, lee-sigmetrics2017, lee-taco2016, lee-pact2015,
      liu-isca2012, liu-isca2013, patel-isca2017, chang-hpca2014,
      seshadri-micro2013, seshadri-micro2017, hassan-hpca2017, kim-cal2015,
      kim-isca2014, kim-hpca2018}} for a detailed background on
  DRAM.} The latencies of these three DRAM operations, as defined by vendor
specifications, have \emph{not} improved significantly in the past 18 years, as
depicted in Figure~\ref{fig:latency_trend}. This is especially true when we
compare \new{latency improvements} to the capacity (128$\times$) and bandwidth
improvements (20$\times$)~\cite{chang-thesis2017} commodity DRAM chips
experienced in the past 18 years. In fact, the activation and precharge
latencies \emph{increased} from 2013 to 2015, when DDR \new{DRAM} transitioned
from the third generation (12.5ns for DDR3-1600J~\cite{jedec-ddr3}) to the
fourth generation (14.06ns for DDR4-2133P~\cite{jedec-ddr4}). As the latencies
specified by vendors have not reduced over time, the memory latency remains as a
critical system performance bottleneck in many modern applications, such as big
data workloads~\cite{clapp-2015} and Google's warehouse-scale
workloads~\cite{kanev-isca2015}.

\figputHSL{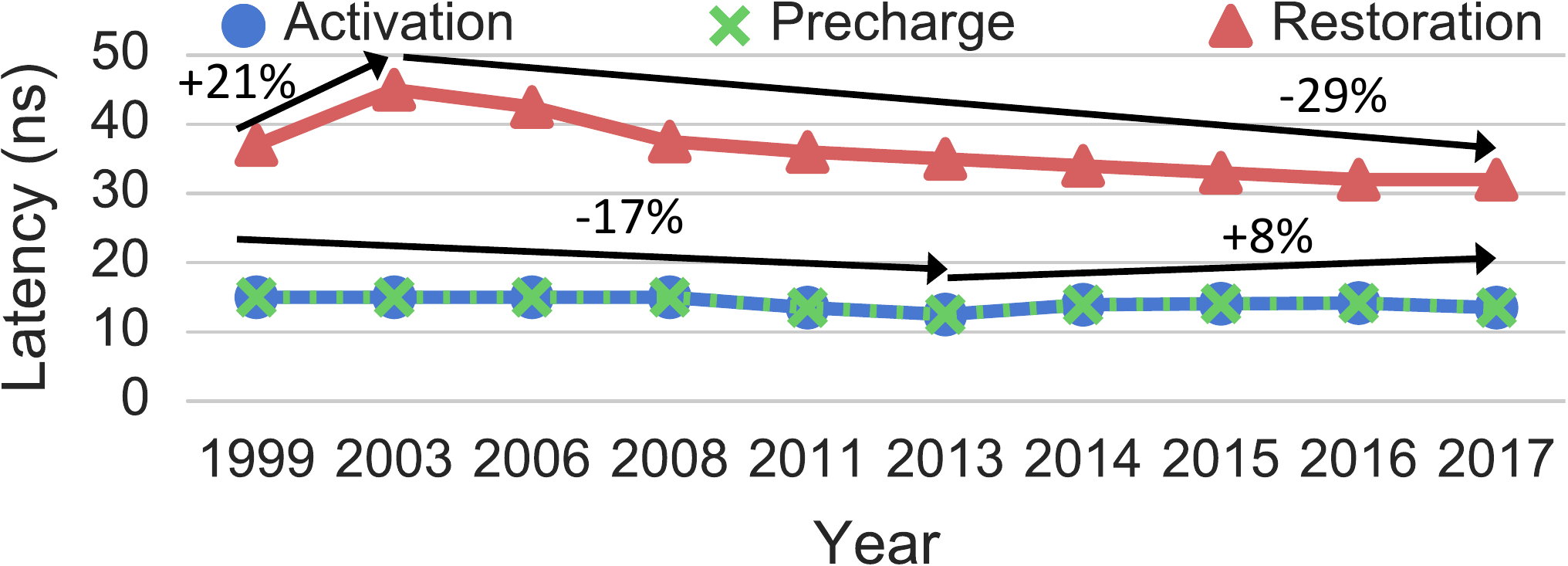}{0.4}{DRAM latency trends over
  time~\cite{micronSDR_128Mb,jedec-ddr2,jedec-ddr3,jedec-ddr4}. \fix{Adapted}
  from~\cite{chang-sigmetrics2016}.}{latency_trend}

\section{Motivation}

In this work, we observe that the \new{three fundamental DRAM} operations can
\emph{actually} complete with a much lower latency for many DRAM cells than the
vendor specification, because \emph{there is inherent latency variation present
  across the DRAM cells within a DRAM chip}.  This is a result of manufacturing
process variation, which causes the \emph{sizes} and \emph{strengths} of cells
to be different, thus making some cells faster and other cells slower to be
accessed reliably\fix{~\cite{li11}}. The speed gap between the fastest and slowest
DRAM cells is getting worse~\cite{kanad_dram_book, pvt_book}, as the technology
node continues to scale down to sub-20nm feature sizes. Unfortunately, instead
of optimizing the latency specifications for the common case, DRAM vendors use a
single set of standard access latencies, called timing parameters, which provide
reliable operation guarantees for the \emph{worst case} (i.e., the slowest
cells), to maximize manufacturing yield.

We experimentally demonstrate that significant latency variation is present
across DRAM cells in 240 DDR3 DRAM chips from three major vendors, and that a
large fraction of cells can be read reliably even if the
activation/restoration/precharge latencies are reduced significantly. By
repeatedly testing these DRAM chips, we observe that access latency variation
exhibits spatial locality within DRAM --- slower cells cluster in certain
regions of a DRAM chip.  In \secref{flydram}, we propose a new mechanism, called
\emph{FLY-DRAM}, which exploits the lower latencies of DRAM regions with faster
cells by introducing heterogeneous timing parameters into the memory controller.
By analyzing and exploiting the latency variation that exists in DRAM cells, we
can greatly reduce the DRAM access latency.

We discuss our major \fix{experimental} observations in
\secref{act_lat_analysis}. For a detailed discussion on all of our observations,
we refer the reader to our SIGMETRICS 2016 paper~\cite{chang-sigmetrics2016}.

\ignore{
Based on the findings from our experimental characterization, we propose and
evaluate a new mechanism, \mech (\mechlong), which exploits the lower latencies
of DRAM regions with faster cells by introducing heterogeneous timing parameters
into the memory controller.  We find that \mech improves performance in an
8-core system by 13.3\%, 17.6\%, and 19.5\%, on average, for each of the three
different vendors' real DRAM chips, across a wide range of applications.
}

% !TEX root=../paper.tex

\section{Latency Variation Analysis}
\label{sec:act_lat_analysis}

To capture the effect of latency variation in modern DDR3 DRAM chips, we tune
the timing parameters that control the amount of time taken for each of the
fundamental DRAM operations. We developed an FPGA-based DRAM testing
platform~\cite{hassan-hpca2017} that allows us to precisely control the timing
parameter values and the tested DRAM location (i.e., banks, rows, and columns).
\fix{A photo of the platform is shown in \figref{dram_fpga_white}.} Using this
platform, we characterize latency variation on a total of 30 DDR3 DRAM modules
(or \emph{DIMMs}), comprising 240 DRAM chips from three major vendors. Each chip
has a 1Gb density. Thus, each of our DIMMs has a 1GB capacity.
\tabref{dimm_list} lists the \new{relevant} information about the tested DRAM
modules. Unless otherwise specified, we test modules at an ambient temperature
of 20$\pm$1\celsius. For results using higher temperatures, we refer the reader
to Section~4.5 of our SIGMETRICS 2016 paper~\cite{chang-sigmetrics2016}.

\figputHS{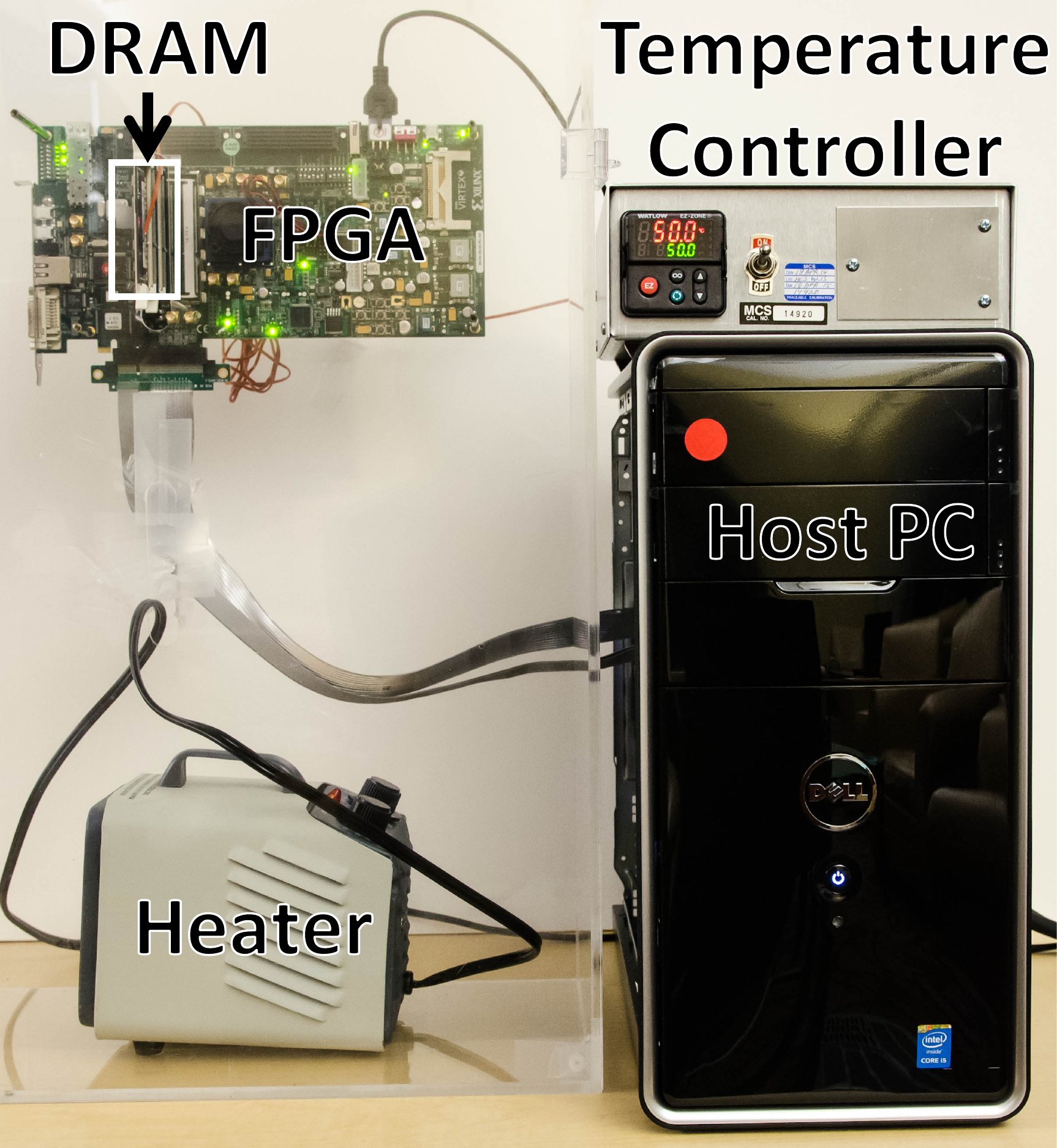}{0.28}{FPGA-based DRAM testing infrastructure.
  Reproduced from \cite{chang-sigmetrics2016}.}

\begin{table}[h]
  \small
  \centering
    \setlength{\tabcolsep}{.35em}

    \begin{tabular}{cccc}
        \toprule
        \multirow{2}{*}{Vendor} & Total Number & Timing (ns) &
        Assembly  \\
        & of Chips & (\trcd/\trp/\tras) & Year  \\
        \midrule
        A (8 DIMMs) & 64 & 13.125/13.125/35-36 & 2012-13\\
        B (9 DIMMs) & 72 & 13.75/13.75/35 & 2011-12 \\
        C (13 DIMMs) & 104 & 13.75/13.75/34-36 & 2011-12 \\
        \bottomrule
    \end{tabular}
  \caption{Main properties of the tested DIMMs. Reproduced from \cite{chang-sigmetrics2016}.}
  \label{tab:dimm_list}
\end{table}

In this section, we present a short summary of our key results on varying the
activation, precharge, and restoration latencies, which are controlled by the
\trcd, \trp, and \tras timing parameters, respectively. For more details on the
experimental results and observations, see Sections~4--6 of our SIGMETRICS 2016
paper~\cite{chang-sigmetrics2016}.

\subsection{Behavior of Timing Errors}

We analyze the variation in the latencies of activation, precharge, and
restoration by operating DRAM at multiple reduced latencies for each of these
operations. Faster cells do \emph{not} get affected by the reduced timings, and
can be accessed reliably without any errors; however, slower cells {\em cannot}
be read reliably with reduced latencies for the three operations, leading to bit
flips. In this work, we define a {\em timing error} as a bit flip in a cell that
occurs due to a reduced-latency access, and characterize timing errors incurred
by the three DRAM operations.

Our experiments yield several \textbf{new observations} on the behavior of
timing errors. When we reduce the three latencies, we observe that each latency
exhibits a different level of impact on the inherently-slower cells. Lowering
the activation latency (\trcd) affects \emph{only} the cells (data) read in the
first accessed cache line, but not the subsequently read cache lines from the
same row. This is mainly due to two reasons.
\fix{First, a \crd command accesses \new{only} its corresponding sense amplifiers, without
accessing the other columns. Hence, a \newt{\crd's effect is isolated to} its
target cache line.}
%First, although an \act is a
%\emph{row-level command},  a \crd is a \emph{column-level} command (i.e., it
%reads only a small portion of the row). Therefore, the impact of reducing \trcd
%can only be observed in the cache line selected by the \crd.
Second, by the time a subsequent \crd is issued to the same activated row, a
sufficient amount of time has already passed for the row buffer to fully sense
and latch in the row data. In contrast, lowering the restoration (\tras) or
precharge (\trp) latencies affects \emph{all} cells in the activated row (see
Section~5 of our SIGMETRICS 2016 paper~\cite{chang-sigmetrics2016} for a detailed explanation).
\fix{Lowering these latencies affects the entire row because these commands
  operate at the row level, and they directly affect the restoration and sensing
  of \emph{all} cells in the row.}

We also find that the number of timing errors introduced is very sensitive to
reducing the activation or precharge latency, but not that sensitive to reducing
the restoration latency. We conclude that different levels of mitigation are
required to address the timing errors that result from lowering each of the
different DRAM operation latencies, and that reducing restoration latency to the
lowest levels allowed by our infrastructure does \emph{not} introduce timing
errors in our experiments \fix{(see Section 6 in our SIGMETRICS 2016
  paper~\cite{chang-sigmetrics2016})}.

\subsection{Timing Error Distribution}

We briefly present the distribution of activation and precharge errors collected
from all of the tests conducted on every DIMM. Figure~\ref{fig:rcdbox} shows the
box plots of the \emph{bit error rate} (BER) observed on every DIMM as
activation latency (\trcd) varies. The BER is defined as the fraction of bits
\fix{with errors} due to reducing \trcd in the total population of tested bits.
In other words, the BER represents the fraction of cells that cannot operate
reliably under the specified shortened latency. The box plot shows the maximum
and minimum BER of all of our tested DIMMs as whiskers, and the box shows the
quartiles of the distribution. In addition, we show \emph{all} observation
points for each specific \trcd/\trp value by overlaying them on top of their
corresponding box plot. Each point shows a BER collected from one round of tests
on one DIMM with a specific data pattern and \trcd value.  For box plots showing
the BER distribution when the precharge latency (\trp) is reduced, see Figure~12
in the original paper~\cite{chang-sigmetrics2016}. We make two observations from
the BER distributions when reducing \trcd or \trp.

\figputHSL{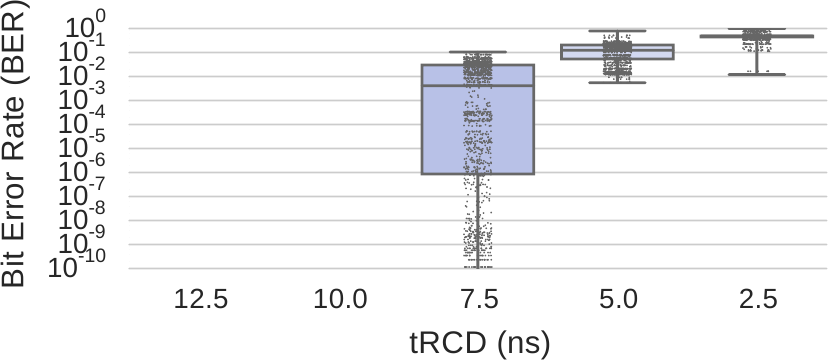}{1}{Bit error rate \new{of
    all DIMMs with reduced \trcd}. Reproduced
  from~\cite{chang-sigmetrics2016}.}{rcdbox}

%\figputHSL{trp/ber/all_ber_plots/all_trp_box_log}{0.8}{Bit error rate
%    of all DIMMs with reduced \trp.}{rpbox}

First, at \trcd or \trp values of 12.5ns and 10ns, we observe \emph{no timing
errors on any DIMM} due to reduced activation or precharge latency. This shows
that the \trcd/\trp latencies of the slowest cells in our tested DIMMs likely
fall between 7.5 and 10ns, which are lower than the value provided in the vendor
specifications (13.125ns).
DRAM vendors use the extra latency as \new{a \emph{guardband}} to provide
additional protection against process variation.

Second, there exists a large BER variation among DIMMs at \trcd of 7.5ns, and
the BER variation becomes smaller as the \trcd or \trp value decreases. The
number of fast cells that can operate at \trcde7.5ns or \trpe7.5ns varies
significantly across different DIMMs. These results demonstrate that there
exists significant latency variation among and within DIMMs, as not all of the
cells exhibit timing errors at 7.5ns.

% , which also demonstrates significant precharge
% latency variation.

%%%%%%%%%%%

\subsection{Spatial Locality of Timing Errors}
\label{ssec:locality}

\begin{figure}[!h]
\centering
\captionsetup[subfigure]{justification=centering}

\subcaptionbox{Activation latency (\trcd) \\at 7.5ns (43\% reduction).\label{fig:rcd_col_err}}[0.48\linewidth] {
  \includegraphics[width=0.5\linewidth, height=80pt]{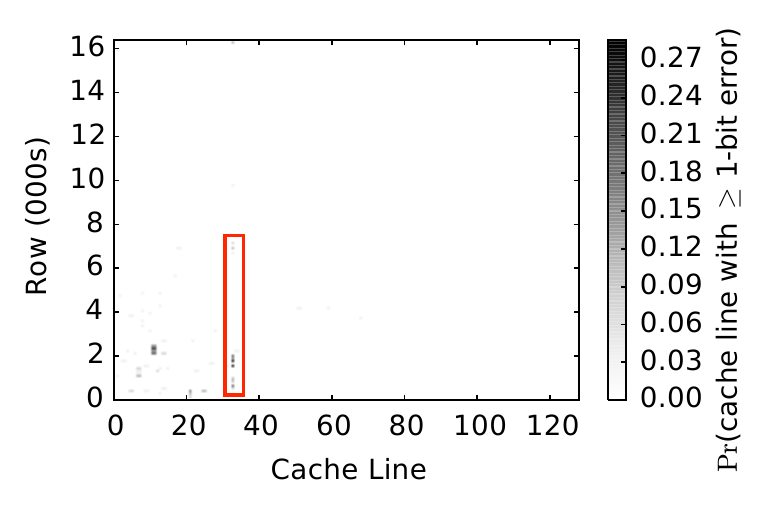}
}
\subcaptionbox{Precharge latency (\trp) \\at 7.5ns (43\% reduction).\label{fig:rp_col_err}}[0.48\linewidth] {
  \includegraphics[width=0.5\linewidth, height=80pt]{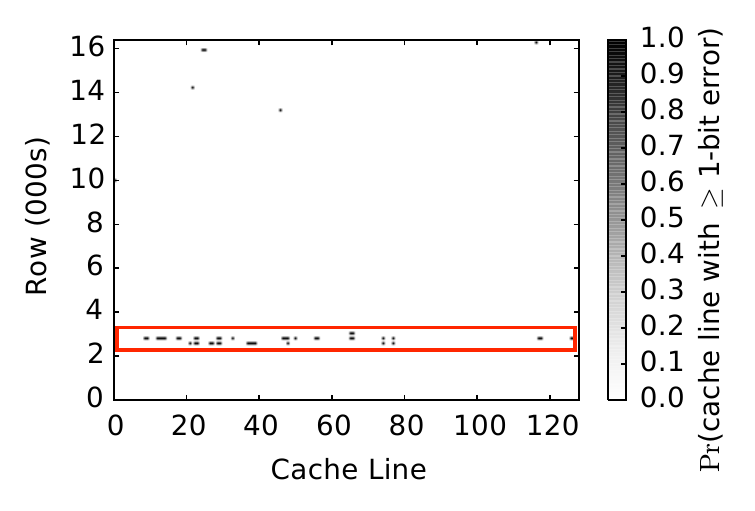}
}
\caption{Probability of observing timing errors in one DIMM. Adapted
  from~\cite{chang-sigmetrics2016}.}
\label{fig:err_loc}
\end{figure}

In this section, we investigate the location and distribution of timing errors \emph{within} a DIMM
when the activation or precharge latencies are reduced. Figure~\ref{fig:err_loc} shows the
probability of every cache line (64B) in one bank of a specific DIMM observing
at least 1 bit of error with reduced activation latency (\figref{rcd_col_err}) or precharge latency
(\figref{rp_col_err}). See \cite{chang-sigmetrics2016} for additional results.
The x-axis and y-axis indicate the cache line number and row number (in
thousands), respectively. In our tested DIMMs, a row size is 8KB,
\new{comprising} 128 cache lines.

The main observation is that timing errors due to reducing activation or
precharge latency are \emph{not} distributed uniformly across locations within
this DIMM. Timing errors tend to cluster at certain regions of cache lines. For
the remaining cache lines, we observe that they do not exhibit timing errors due
to reduced latency throughout the experiments. We observe similar
characteristics in other DIMMs --- timing errors concentrate within certain
spatial regions of memory.

We hypothesize that the cause of \new{the spatial locality of timing errors is
due to the locality of} variation in the fabrication process during
manufacturing. Certain cache line locations can end up with less robust
components, such as weaker sense amplifiers, weaker cells, or higher resistance
bitlines.

%%%%%%

\subsection{Other Characterization Results}

We briefly summarize our other observations on the effects of reducing timing
parameters. First, we analyze the number of timing errors that occur when DRAM
access latencies are reduced, and experimentally demonstrate that most of the
erroneous cache lines have a single-bit error, with only a small fraction of
cache lines experiencing more than one bit flip (\new{see Section~4.7
of our SIGMETRICS 2016 paper}~\cite{chang-sigmetrics2016}). We conclude, therefore, that using simple
error-correcting codes (ECC) can correct \emph{most} of these errors, thereby
enabling lower latency for many inherently slower cells \new{(see Section~4.8 of
our SIGMETRICS 2016 paper~\cite{chang-sigmetrics2016} for a detailed analysis
of ECC)}.

Second, we find that the stored data pattern in cells affects access latency
variation. Certain patterns lead to more timing errors than others. For example,
the bit value \texttt{1} can be read significantly more reliably at a reduced
access latency than the bit value \texttt{0} (\new{see Section~4.4
of our SIGMETRICS 2016 paper}~\cite{chang-sigmetrics2016}). \fix{This observation is similar to the data
  pattern dependence observation made for retention times of DRAM
  cells~\cite{liu-isca2013,khan-sigmetrics2014,khan-dsn2016,khan-cal2016,khan-micro2017,patel-isca2017}.}

Third, we find no clear correlation between temperature and variation in cell
access latency. We believe that it is not essential for latency reduction
techniques that exploit such variation to be aware of the operating temperature
(Section~4.5 in~\cite{chang-sigmetrics2016}).

% !TEX root=../paper.tex

\section{Exploiting Latency Variation}
\label{sec:flydram}

Based on our extensive experimental characterization and \fix{new observations
on latency-reliability trade-offs in modern DRAM chips}, we propose a new
hardware mechanism, called \emph{\mechlong} (\mech), to reduce DRAM latency for
better system performance. \mech exploits the key observation that 
{\em (i)}~different cells \fix{can operate reliably at different DRAM latencies,} and
{\em (ii)}~there is a strong correlation between the location of a cell and the
lowest latency that the cell can \fix{operate reliably at}. The key idea of
\mech is to {\em (i)}~categorize the DRAM cells into fast and slow regions, {\em
  (ii)}~expose this categorization to the memory controller, and 
{\em (iii)}~reduce overall DRAM latency by accessing the fast regions with a lower latency.

\newt{The} \mech memory controller \romnum{i} loads the latency profiling
results~\cite{chang-sigmetrics2016} into on-chip SRAM at system boot time,
\romnum{ii} looks up the profiled latency for each memory \newt{request} based
on its memory address, and \romnum{iii} applies the \newt{corresponding latency
  to the request}. By reducing the values of \trcd, \tras, and \trp for some
memory requests, \mech improves overall system performance. In addition, we also
propose an OS page allocator design that exploits the latency variation in DRAM
to improve system performance (see Section~7.2 of our
paper~\cite{chang-sigmetrics2016}).

There are two key design challenges of \mech. The first challenge is determining
the fraction of fast cells within a DRAM chip and the innate access latency of
the fast cells. Since DRAM vendors have detailed information on their DRAM chips
from the DRAM post-production tests, DRAM vendors can embed the \fix{latency}
profiling results in the \fix{Serial} Presence Detect (SPD) circuitry (a ROM present
in each DIMM)~\cite{jedec-spd}. The memory controller can read the profiling
results from the SPD circuitry during DRAM initialization, and apply the correct
latency for each DRAM region.

The second design challenge is limiting the storage overhead of the latency
profiling results. Recording the shortest latency for each cache line can incur
a large storage overhead. Fortunately, the storage overhead can be reduced based
on \newt{a new} observation of ours. As discussed in \ssecref{locality}, timing
errors typically concentrate at certain DRAM regions. Therefore, \mech records
the shortest latency at the granularity of DRAM regions (i.e., a group of
adjacent cache lines, \fix{rows, or banks}). One can imagine using more
sophisticated structures, such as Bloom Filters~\cite{bloom-cacm70}, to provide
finer-grained latency information within a reasonable storage overhead, as shown
in prior work on variable DRAM refresh \fix{intervals}~\cite{liu-isca2012,qureshi-dsn2015}.

\subsection{Summary of Results}

We evaluate \mech on on an 8-core system with a wide variety of workloads by
using Ramulator~\cite{safari-github, kim-cal2015}, a cycle-level
\fix{open-source} DRAM simulator \fix{developed by our research group}.
Table~\ref{tab:sys-config} summarizes the configuration of our evaluated system.
We use the standard DDR3-1333H timing parameters~\cite{jedec-ddr3} as our
baseline.

%%%%%%%%%%%%%%%%%
% Single Column
%%%%%%%%%%%%%%%%%
\begin{table}[h]
%\footnotesize
%\scriptsize
    \renewcommand{\arraystretch}{0.9}
    \small
  \centering
    \setlength{\tabcolsep}{.45em}
    \begin{tabular}{ll}
        \toprule
        \textbf{Processor}   & 8 cores, 3.3 GHz, OoO 128-entry window \\
                                     %& 4-wide issue, 16 MSHRs per core \\
        \midrule

        \textbf{LLC} & 8 MB shared, 8-way set associative \\

        \midrule

        %Memory Controller & 64/64-entry read/write request queue, FR-FCFS~\cite{rixner-isca2000} \\
        %\midrule

        \multirow{3}{*}{\textbf{DRAM}} & DDR3-1333H~\cite{jedec-ddr3},
        open-row policy~\cite{rixner-isca2000,kim-hpca2010,kim-micro2010}, \\
        & 2 channels,  1 rank per channel, 8 banks per rank, \\
        & Baseline: \trcd/\tcl/\trp= 13.125ns, \tras= 36ns \\

        %DRAM & DDR3-1333H~\cite{micronDDR3_4Gb}, 2 channels, 1 rank and 8 banks \\
        %\midrule

        %DRAM  & Baseline: \trcd/\tcl/\trp= 12, \tras= 36 \\
        %Timing (ns) & VL-DRAM: \trcd/\trp= 7.5, \tcl= 12, \tras= 27 \\

        \bottomrule
    \end{tabular}
  \caption{Evaluated system configuration. Adapted from
    \cite{chang-sigmetrics2016}.}
  \label{tab:sys-config}%
  %\vspace{-0.15in}
\end{table}

\begin{figure}[h]
  \centering
  \includegraphics[scale=1]{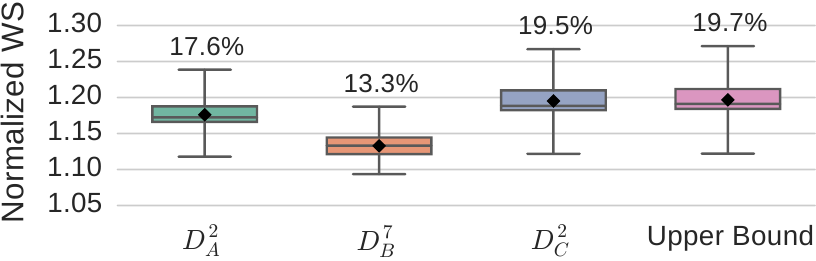}
  \caption{System performance \newt{improvement} of \mech for various DIMMs. Reproduced from~\cite{chang-sigmetrics2016}.\label{fig:vldram}}
\end{figure}

\figref{vldram} illustrates the system performance improvement of \mech over the
baseline (DDR3-1333) for 40 workloads. The x-axis indicates each of the
evaluated DRAM configurations. \dimm{2}{A}, \dimm{7}{B}, and \dimm{2}{C}
correspond to latency profiles collected from three real DIMMs. Our SIGMETRICS
2016 paper~\cite{chang-sigmetrics2016} describes these real-DRAM profiles in
more detail.

For these three DIMMs, \mech improves system performance significantly, by
17.6\%, 13.3\%, and 19.5\% on average across all 40 workloads. This is because
\mech reduces the latency of \trcd, \trp, and \tras by 42.8\%, 42.8\%, and 25\%,
respectively, for \fix{a large fraction of} cache lines. In particular, DIMM \dimm{2}{C}, which has a
99\% of cells that operate reliably at low \trcd and \trp, performs within 1\%
of the upper-bound performance (19.7\% on average), \fix{which is obtained by
  operating \ch{all} DRAM cells at low \trcd and \trp}. We \fix{conclude} that \mech is an
effective mechanism to improve system performance by exploiting the widespread
latency variation present across DRAM cells.

\fix{As we show in our SIGMETRICS 2016 paper~\cite{chang-sigmetrics2016},
  \mbox{FLY-DRAM} can take advantage of an intelligent \mbox{DRAM-aware} page
  allocator that allocates frequently used and latency-critical pages in fast
  DRAM regions. We leave the detailed design and evaluation of such an allocator
  to future work.}

% !TEX root=../paper.tex

\section{Related Work}

To our knowledge, this is the first work to {\em (i)}~provide a detailed
experimental characterization and analysis of latency variation for three
\newt{major} DRAM operations (\trcd, \trp, and \tras) \emph{across different
  cells within a DRAM chip}, {\em (ii)}~demonstrate that a reduction in latency
for each of these fundamental operations has a different impact on different
cells, {\em (iii)}~show that access latency variation exhibits spatial locality,
{\em (iv)}~demonstrate that the error rate \newt{due to reduced latencies} is
correlated with the \newt{stored} data pattern but \emph{not} \newthree{conclusively
  correlated} with temperature, and {\em (v)}~propose mechanisms that take
advantage of variation \emph{within a DRAM chip} to improve system performance.
We discuss the most closely related works here.

\subsection{DRAM Latency Variation} Adaptive-Latency DRAM (AL-DRAM) also
characterizes and exploits DRAM latency variation, but does so at a much coarser
granularity~\cite{lee-hpca2015}. This work experimentally characterizes latency
variation \emph{across} different DRAM chips under \emph{different} operating
temperatures. AL-DRAM sets a uniform operation latency for the \emph{entire}
DIMM. In contrast, our work characterizes latency variation \emph{within each
  chip}, at the granularity of individual DRAM cells. Our mechanism, \mech, can
be combined with AL-DRAM to further improve performance.\ch{\footnote{\ch{A
description of the AL-DRAM work and its impact is provided in a companion
article in the very same issue of this journal~\cite{aldram.tar18}.}}}

A recent work by Lee et al.~\cite{lee-sigmetrics2017} also
observes latency variation within DRAM chips. The work analyzes the variation
that is due to the circuit design of DRAM components, which it calls
\emph{design-induced variation}. Furthermore, it proposes a new profiling
technique to identify the lowest DRAM latency without introducing errors. In
this work, we provide the \emph{first} detailed experimental
characterization and analysis of the general latency variation phenomenon within
real DRAM chips. Our analysis is broad and is not limited to design-induced
variation. Our proposal of exploiting latency variation, FLY-DRAM can employ Lee
et al.'s new profiling mechanism~\cite{lee-sigmetrics2017} to
identify additional latency variation regions for reducing access latency.

Chandrasekar et al.\ study the potential of reducing some DRAM timing
parameters~\cite{chandrasekar-date2014}. Similar to AL-DRAM, this work observes
and characterizes latency variation \emph{across} DIMMs, whereas our work
studies variation \newt{across} cells \emph{within a DRAM chip}.

\subsection{DRAM Error Studies} There are several studies that characterize
various errors in DRAM. Many of these works observe how specific factors affect
DRAM errors, analyzing the impact of
temperature~\cite{el-sayed-sigmetrics2012,lee-hpca2015} and hard
errors~\cite{hwang-asplos2012}. Other works have conducted studies of DRAM error
rates in the field, studying failures across a large sample
size~\cite{schroeder-sigmetrics2009, meza-dsn2015, li-usenixatc2010,
  sridharan-sc2012, sridharan-asplos2015}. There are also works that have
studied errors through controlled experiments, investigating errors due to
retention time~\fix{\cite{khan-cal2016,khan-micro2017,liu-isca2013,
    khan-sigmetrics2014, khan-dsn2016, qureshi-dsn2015, patel-isca2017,
    hassan-hpca2017}}, disturbance from neighboring DRAM
cells~\cite{mutlu-date2017,kim-isca2014}, latency variation across/within DRAM
chips~\cite{lee-thesis2016,chandrasekar-date2014, lee-hpca2015,
  lee-sigmetrics2017}, and supply voltage~\cite{chang-sigmetrics2017}. None of
these works study errors due to latency variation across the cells \emph{within}
a DRAM chip, which we extensively characterize in our work.

\subsection{DRAM Latency Reduction} Several types of commodity DRAM (Micron's
RLDRAM~\cite{micron-rldram3} and Fujitsu's FCRAM~\cite{sato-vlsic1998}) provide
low latency at the cost of high area overhead~\cite{kim-isca2012, lee-hpca2013}.
Many prior works
\fix{(e.g.,~\cite{seshadri-micro2017,mutlu-date2017,mutlu-imw2013,chang-hpca2014,kim-isca2012,
    lee-hpca2013,o-isca2014,chang-hpca2016,hidaka-ieeemicro90,
    seshadri-micro2013, seshadri-cal2015,
    zhang-isca2014,lu-micro2015,son-isca2013})} propose various architectural
changes \emph{within} DRAM chips to reduce latency. In contrast, \mech does not
require any changes to a DRAM chip. Other
works~\newt{\cite{shin-hpca2014,lee-tech2010,hassan-hpca2016,seshadri-isca2014,seshadri-micro2015}}
reduce DRAM latency by changing the memory controller, and \mech is
complementary to them.

\subsection{ECC DRAM} Many memory systems incorporate ECC DIMMs, which store
information used to correct data during a read operation. Prior work
(e.g.,~\cite{vecc, yoon.isca12.boom, udipi2012lot, kim2015bamboo, li2012mage,
  khan-sigmetrics2014, wilkerson-isca2010, gongclean, jian2013low}) proposes
more flexible or more powerful ECC schemes for DRAM. While these ECC mechanisms
are designed to protect against faults using standard DRAM timings, we show that
they also have the potential to correct timing errors that occur due to reduced
DRAM latencies. \fix{A recent work by Lee et al.~\cite{lee-sigmetrics2017}
  exploits this observation and uses ECC to correct errors that occur due to
  reduced latency in DRAM.}

% Although these ECC mechanisms are mainly used to protect against hardware
% faults in DRAM, in this work, we show that ECC DIMM has the potential to
% provide low latency and reliability.

\subsection{Other Latency Reduction Mechanisms}
  %\subsection{Reducing DRAM Latency by Processing in Memory}

\fix{ Various prior
  works~\cite{ahn-isca2015,ahn-isca2015-2,7056040,7429299,guo-wondp14,592312,
    seshadri-cal2015,mai-isca2000,draper-ics2002,
    seshadri-micro2015,hsieh-iccd2016,hsieh-isca2016,
    amirali-cal2016, stone-1970, fraguela-2003,375174,808425,
    4115697,694774,sura-2015,zhang-2014,akin-isca2015,
    babarinsa-2015,7446059,6844483,pattnaik-pact2016, seshadri-thesis2016,
    seshadri-micro2017, chang-hpca2016,kim-apbc2018,hashemi-isca2016, ami-asplos2018} examine
  processing in memory to reduce DRAM latency. Other prior works propose memory
  scheduling techniques,\ch{~\cite{kim-hpca2010, muralidhara-micro2011, lee-micro2008,
  kim-micro2010, subramanian-tpds2016, subramanian-hpca2013, subramanian-micro2015,
    subramanian-iccd2014, mutlu-isca2008, mutlu-micro2007, usui-taco2016, ghose-isca2013,
    ausavarungnirun-isca2012, moscibroda-usenix2007, ipek-isca2008}},
     which generally reduce latency to access
  DRAM. Our analyses and
  techniques can be combined with these works to enable \ch{further low-latency operation}. }

%\new{Many prior works propose memory scheduling techniques, which generally
%reduce latency to access DRAM~\cite{kim-hpca2010, kim-micro2010,
%  subramanian-tpds2016, subramanian-iccd2014, mutlu-isca2008, mutlu-micro2007,
%  usui-taco2016, ausavarungnirun-isca2012, moscibroda-usenix2007}. Other works
%propose mechanisms to perform in-memory computation to reduce data movement and
%access latency~\cite{hsieh-iccd2016, seshadri-micro2017, amirali-cal2016,
%  hsieh-isca2016}. LISA is complementary to these works, and it can work
%synergistically with in-memory computation mechanisms by enabling fast
%aggregation of data.}

% !TEX root=../paper.tex
\section{Significance}

\fix{Our SIGMETRICS 2016 paper~\cite{chang-sigmetrics2016} presents a new 
\ch{experimental characterization and analysis of} latency variation in
modern DRAM chips. In this section, we describe the potential impact that our
study can have on the research community and industry.}

\subsection{Potential Research Impact}

Our paper develops a new way of using manufactured DRAM chips: accessing
different regions of memory using each region's inherent latency instead of a
homogeneous fixed standard latency for all regions of memory. \fix{We show that {\em
  (i)} there is significant latency variation within a DRAM chip, and {\em (ii)}
it is possible to exploit the variation with simple mechanisms.}
We believe one key impact of our paper is demonstrating the effectiveness
of designing memory optimizations based on real-world characterization. We
expect that this same principle can be used to craft new memory architectures
for both existing and future memory technologies, such as SRAM,
\fix{PCM~\cite{lee-isca2009, lee-ieeemicro2010, qureshi-isca2009,
    yoon-taco2014,lee-cacm2010,qureshi-micro2009,yoon-iccd2012},
  STT-MRAM~\cite{ku-ispass2013,guo-isca2009,chang-hpca2013}, or
  RRAM~\cite{wong-ieee2012}. }
% We
% believe the key impact of our paper will come from the basic principle of the
% characterize-and-optimize approach toward memory devices. This principle is
% likely applicable on using other types of existing or future memory
% technologies, such as SRAM, STT-RAM, PCM, etc.

Our work exposes several opportunities for both operating systems and hardware
to further optimize for memory access latency. We have open-sourced our raw
characterization data, to allow other researchers to further analyze and build
off of our work~\cite{safari-github}. Other researchers can find many other ways
to take advantage of the insights and the characterization data we provide. Our
FLY-DRAM implementation is also available as part of \fix{the open-source
  release of} Ramulator~\cite{kim-cal2015,ram-github}.

\textbf{ECC to Reduce Latency.}
%\noindent\textbf{5.1.1 ECC to Reduce Latency.}
In our paper, we analyze the distribution of timing errors (due to reduced
latency) at the granularity of data beats,  as conventional error-correcting
codes (ECC) work at the same granularity. Our data shows that many of the
erroneous data beats experience only a single-bit error, while the majority of
the data beats contain no errors. Therefore, this creates an opportunity for
applying ECC to correct timing errors. We also envision an opportunity for
applying ECC to only certain regions of DRAM, which takes advantage of the
spatial locality of timing errors exposed by our work. \fix{Lee et
  al.~\cite{lee-sigmetrics2017} provide examples of the use of ECC to reduce
  latency further, but they apply ECC globally to the entire DRAM chip. We
  believe a significant opportunity exists in customizing ECC to latency errors
  and different DRAM \ch{reliability} issues.}

\textbf{Data Pattern Dependence.}
We find that timing errors caused by reducing activation latency are dependent
on the stored data pattern. Reading bit \texttt{1} is significantly more
reliable than bit \texttt{0} at reduced activation latencies. This asymmetric
sensing strength can potentially be a good direction for studying DRAM
reliability. Currently, DRAM commonly employs data bus
inversion~\cite{jedec-ddr4} as an encoding scheme to reduce toggle rate on the
data bus, thereby saving channel power\fix{~\cite{gena-hpca2016}}. Similar
encoding techniques can be developed to reduce bit \texttt{0}s and increase the
overall number of \texttt{1}s in data. \fix{We believe that} developing
asymmetric data encodings or ECC mechanisms that favor \texttt{1}s over
\texttt{0}s is a promising research direction to improve DRAM reliability.

\textbf{DRAM-Aware Page Allocator.}
We developed a hardware mechanism (\mech) that exploits latency variation to
improve system performance in a software-transparent manner.  Researchers can
take better advantage of the variation by exposing the different latency regions
to the software stack. In our \fix{SIGMETRICS 2016}
paper~\cite{chang-sigmetrics2016}, we discuss the potential of a DRAM-aware page
allocator in the OS (Section 7.2), which can improve FLY-DRAM performance by
intelligently mapping more frequently-accessed application pages to faster DRAM
regions. We believe that the key idea of enabling the OS to allocate pages based
on the accessed memory region's latency can be applied to other types of memory
characteristics (e.g., energy efficiency \fix{or
  voltage~\cite{chang-sigmetrics2017,david-icac2011}}) without needing to modify the
architecture.

\textbf{Applicability to Other Memory Technologies.}
\fix{In this work, we focus on characterizing only DRAM technology. A class of
emerging memory technology is non-volatile memory (NVM), which has the
capability of retaining data even when the memory is not powered. Since the memory organization
of NVM mostly resembles that of DRAM\ch{~\cite{meza-iccd2012, yoon-iccd2012,
lee-isca2009}}, we believe that our characterization and
optimization can be extended to different types of NVMs, such as
PCM~\cite{lee-isca2009, lee-ieeemicro2010, qureshi-isca2009,
  yoon-taco2014,lee-cacm2010,qureshi-micro2009,yoon-iccd2012},
STT-MRAM~\cite{ku-ispass2013,guo-isca2009,chang-hpca2013}, 
\ch{or NAND flash memory~\cite{cai-iccd2012,cai-hpca2015,cai-itj2013,luo-msst2015,
luo-hpca2018, cai-dsn2015, cai-hpca2017, cai-date2013,
cai-date2012, cai-ieee2017, cai-ieeearxiv2017, cai-bookchapter2017,
cai-sigmetrics2014, luo-jsac2016}}
to further enhance
their reliability or performance.}

\subsection{Long-Term Impact on Industry}

High main memory latency remains a problem for many modern applications, such as
in-memory databases (e.g., \fix{Redis~\cite{redis}, MemSQL~\cite{memsql},
TimesTen~\cite{timesten}), Spark, Google's datacenter
workloads~\cite{kanev-isca2015,clapp-2015}, and many mobile and interactive
workloads}. We propose two simple ideas that
exploit latency variation in existing DRAM chips. Both can be adopted relatively
easily in the processor architecture (i.e., the memory controller) or in the OS.

In addition to improving memory access latency, reducing the latency of the
three fundamental DRAM operations also increases the effective memory bandwidth.
To fully utilize the available memory bandwidth, memory controllers would have
to maximize the number of \crd or \cwr commands. However, due to interference
between access streams within and across applications, memory controllers need
to constantly open and close rows by issuing \act and \cpre commands \fix{due to
  an increasing number of bank conflicts~\cite{kim-isca2012, hassan-hpca2016}}.
These commands increase the queuing latency of accesses (\crd and \cwr), thus
decreasing the effective memory bandwidth utilization.

As pin count is limited and increasing bus frequency is becoming more difficult
(due to signal integrity issues~\cite{david-icac2011}), our work offers a new
alternative to help improve bandwidth utilization. By reducing the latency of
DRAM operations, which fall on the critical path of DRAM access time, more
accesses per second are allowed, thereby improving the overall effective
bandwidth. Furthermore, improving latency and effective bandwidth also leads to
lower memory energy consumption due to reduced execution time and fewer active
cycles.

\fix{All these benefits (e.g., reduced latency, increased bandwidth, and reduced
  energy) will become much more important as applications become more
  data-intensive and systems become more energy-constrained in the foreseeable
  future\ch{~\cite{mutlu-imw2013, superfri}}.}

In conclusion, we believe that in the longer term, the idea of leveraging
variation in different characteristics (e.g., latency, reliability) inside
memory chips will become more beneficial for both the software and hardware
industry. For example, by making CPU aware of variation behavior in memory
devices, memory vendors have an incentive to sell memory with larger variation
at a lower price, allowing system designers to lower costs with a small amount
of additional logic in hardware. \fix{Many other opportunities to improve system
performance, energy, and cost abound, which we hope the future works can build
upon and exploit.}

\section{Conclusion}

This paper provides the first experimental study that comprehensively
characterizes and analyzes the latency variation within modern DRAM chips for
three fundamental DRAM operations (activation, precharge, and restoration).  We
find that significant latency variation is present across DRAM cells in all 240
of our tested DRAM chips, and that a large fraction of cache lines can be read
reliably even if the activation/restoration/precharge latencies are reduced
significantly. Consequently, exploiting the latency variation in DRAM cells can
greatly reduce the DRAM access latency.
%%We observe that there
%%is a locality in slower cells, clustered in certain regions of a DRAM
%%chip, that can be exploited to develop low-cost mechanisms to
%%reduce latency, where fast regions have lower access latency.
%%
Based on the findings from our experimental characterization, we propose and
evaluate a new mechanism, \mech (\mechlong), which reduces DRAM latency by
exploiting the inherent latency variation in DRAM cells.  \mech reduces DRAM
latency by categorizing the DRAM cells into fast and slow regions, and accessing
the fast regions with a reduced latency.  We demonstrate that \mech can greatly
reduce DRAM latency, leading to significant system performance
\newt{improvements} on a variety of workloads.

We conclude that it is promising to understand and exploit the inherent latency
variation within modern DRAM chips. We hope that the experimental
characterization, analysis, and optimization techniques presented in this paper
will enable the development of other new mechanisms that exploit the latency
variation within DRAM to improve system performance and perhaps reliability.

%
% the
%latency variation in three major DRAM operations
%What
%why
%how
%potential apps
%limitations (couldn't do it on a computer system with memtest)
%Possible avenue for future research
%
%The present study raises the possibility that a link may existing between soft
%errors and timing errors.
%

\section*{Acknowledgments}

We thank \ch{the anonymous reviewers and
SAFARI group members for their} feedback. We acknowledge the support of
Google, Intel, NVIDIA, and Samsung. This research was supported in
part by the ISTC-CC, SRC, and NSF (grants 1212962 and 1320531). Kevin
Chang was supported in part by the SRCEA/Intel Fellowship.

{
\bstctlcite{bstctl:etal, bstctl:nodash, bstctl:simpurl}
\bibliographystyle{IEEEtranS}
\balance
\bibliography{kevin_paper}
}

\end{document}